\documentclass[letterpaper,italian,english,aip,rsi,preprint]{revtex4-1}
\usepackage[T1]{fontenc}
\usepackage[latin9]{inputenc}
\usepackage{units}
\usepackage{amsmath}
\usepackage{amssymb}
\usepackage{graphicx}
\usepackage{esint}

\makeatletter

\pdfpageheight\paperheight
\pdfpagewidth\paperwidth

 \usepackage{fixme}
 \usepackage{acronym}

\usepackage{textcomp}

\makeatother

\usepackage{babel}
\begin{document}

\title{(Almost-)blind locking algorithm for high finesse suspended optical
cavities}

\author{G.~Cella}

\email{giancarlo.cella@pi.infn.it}

\affiliation{Istituto Nazionale di Fisica Nucleare, Largo B. Pontecorvo 3, 56127
Pisa, Italy}

\author{M.~Marchiò}

\email{manuelmrch@gmail.com}

\affiliation{Department of Astronomy, University of Tokyo, 7-3-1 Hongo, Bunkyo-ku,
Tokyo, 113-0033, JAPAN}

\affiliation{National Astronomical Observatory of Japan, 2-21-1 Osawa, Mitaka,
Tokyo, 152-8550, JAPAN}

\date{\today}
\begin{abstract}
Suspended resonant optical cavities are basic building blocks for
several experimental devices. An important issue is the control strategy
required to bring them in the resonant or slightly detuned configuration
needed for their operation, the so-called \emph{locking} procedure.
This can be obtained with a feedback strategy, but the error signal
needed is typically available only when the cavity is near the resonance
with a precision of the order $\Delta L\simeq\lambda_{\ell}{\cal F}^{-1}$,
where ${\cal F}$ is the cavity finesse and $\lambda_{\ell}$ is the
laser's wavelength. When the mirrors are freely swinging the locking
can be attempted only in the short time windows when this condition
is verified. Typically this means that the procedure must be repeated
several times, and that large forces must be applied. In this paper
we describe a different strategy, which tries to take advantage by
the fact that the dynamics of the mirrors is known at least in an
approximate way. We argue that the locking procedure considered can
be more efficient compared with the naive one, with a reduced needed
maximal feedback. Finally we discuss possible generalizations and
we point to future investigations.
\end{abstract}


\maketitle
\selectlanguage{italian}%
\global\long\def\diff{\mathop{}\!\mathrm{d}}

\global\long\def\Diff#1{\mathop{}\!\mathrm{d^{#1}}}

\selectlanguage{english}%

\section{\label{sec:Introduction}Introduction}

Optical resonant cavities provide very effective ways to measure very
small displacements, and are a key component of several experimental
high precision devices, in particular interferometric gravitational
wave detectors such as Virgo Advanced~\citep{0264-9381-32-2-024001}
and LIGO Advanced~\citep{0264-9381-32-7-074001}, which are currently
near the start of new scientific runs with improved sensitivity, or
KAGRA~\citep{PhysRevD.88.043007} which will follow in the near future.
Other examples can be the macroscopic optomechanical devices proposed
by several groups~\citep{PhysRevA.73.023801} for the production
of squeezed states~\citep{Schnabel2004} using ponderomotive effects.

In both cases the cavity mirrors must be able to move along the cavity
axis, for gravitational wave detectors must couple efficiently to
the tidal force of a gravitational wave and ponderomotive devices
must respond to the radiation pressure force. This can be obtained
by suspending the mirrors to a chain of pendula which allow for the
longitudinal motion with the additional bonus of attenuating the seismic
noise above some cut-off frequency $f_{s}$~\citep{0264-9381-19-7-353}.

However seismic attenuation is not effective below $f_{s}$, and the
residual mirror motion could be too large to be compatible with a
resonant cavity working in the linear regime: typically, for a $f_{s}$
of few Hertz the residual mean square displacement is of the order
of $10^{-6}\unit{m}$\citep{peterson}.

The cavity length must be stabilized with a control strategy operating
below the cut off frequency but ineffective above it, where we need
free motion. There are two problems that must be solved. The first
is how to \emph{lock} a cavity which is initially moving in a completely
nonlinear regime, in a sense that will be precised later, with typical
displacements for the mirrors of the order of few laser wavelengths.
The second is how to maintain it in this linear (\emph{locked}) regime
once this is reached. In this paper we will deal only with the lock
acquirement (\emph{locking}) issue.

In order to design a feedback control we need an error signal. This
is provided by the phase shift experimented by a coherent light beam
which enters the cavity and is reflected or transmitted. Let us consider
a resonant cavity of length $L$ built with two mirrors with reflection
coefficient $r_{1,2}$ and transmission coefficient $t_{1,2}$. We
are especially interested in the regime where both the reflectivities
are near one. An important parameter is the cavity finesse ${\cal F}$,
which is defined as the ratio between the full width at half-maximum
bandwidth of its resonances and its free spectral range and can be
written as 
\begin{equation}
{\cal F}=\frac{\pi\sqrt{r_{1}r_{2}}}{1-r_{1}r_{2}}\simeq\frac{\pi}{1-r_{1}r_{2}}
\end{equation}
As we will see in detail later it is possible to measure accurately
the phase shift induced by the cavity on the reflected or transmitted
light only near its resonance peak. For a suspended cavity which is
freely moving the fraction of the time available for the measurement
is thus given by ${\cal F}^{-1}$, and the typical duration of each
passage in the measurement region will be of the order
\begin{equation}
\tau\simeq\frac{\lambda_{\ell}}{v{\cal F}}
\end{equation}
where $v$ is the typical longitudinal speed of the mirrors, which
depends mainly on seismic noise, and $\lambda_{\ell}$ is the wavelength
of the laser. If we want to stop the cavity during this time we need
to apply a force such that 
\begin{equation}
F_{max}\simeq\frac{mv}{\tau}=\frac{mv^{2}}{\lambda_{\ell}}{\cal F}
\end{equation}
which becomes large for an high finesse cavity. Another point to consider
is that the frequency bandwidth $\Delta f$ of a feedback that must
be present only during the measurement time $\tau$ must satisfy
\begin{equation}
\Delta f\geq\frac{1}{4\pi\tau}=\frac{v{\cal F}}{8\pi\lambda_{\ell}}
\end{equation}
In order to be able to design a simple and robust feedback the acceptable
bandwidth must not extend too much, entering in a region where a large
number of internal resonances of the mechanical system are present.
Once again we can meet a problem when the finesse (or the speed $v$)
becomes large.

Our proposal for a different approach to the problem starts from the
consideration that by using measured data about the state of the system,
only when they are available, we neglect a large amount of information
which is known \emph{a priori}. In a general framework we could assume
that the state of the system is described by a finite set of variables
that we will denote collectively with $\boldsymbol{x}$. We do not
measure directly the state, but a different set of parameters $\boldsymbol{y}$
which are connected to the state in a probabilistic way, namely by
a conditional probability$P\left(\boldsymbol{y}\mid\boldsymbol{x}\right)$
whose form is supposed to be known. We suppose also that we know the
conditional probability of $\boldsymbol{x}$ at the time $t$ given
$\boldsymbol{x}^{\prime}$ for $t^{\prime}<t$, which we will denote
$P\left(\boldsymbol{x},t\mid\boldsymbol{x}^{\prime},t^{\prime}\right)$,
and the initial \emph{a priori} probability for the state $P(\boldsymbol{x})$.

Our first aim is the estimation of the probability $P\left(\boldsymbol{x},t\mid\boldsymbol{y}_{1},t_{1};\cdots\boldsymbol{y}_{n},t_{n}\right)$
where $t>t_{n}$, $\boldsymbol{y}_{k}$ is a previous measurement
done at the time $t_{k}<t_{k+1}$ and $0<k<n$. Formally a simple
recursive formula can be written for it using a \emph{prediction step
}which uses the knowledge about the dynamic of the system

\begin{widetext}

\begin{equation}
P\left(\boldsymbol{x},t\mid\boldsymbol{y}_{1},t_{1};\cdots;\boldsymbol{y}_{n},t_{n}\right)=\int P\left(\boldsymbol{x},t\mid\boldsymbol{x}_{n},t_{n}\right)P\left(\boldsymbol{x}_{n},t_{n}\mid\boldsymbol{y}_{1},t_{1};\cdots;\boldsymbol{y}_{n},t_{n}\right)\diff\boldsymbol{x}_{n}\label{eq:predictionSTP}
\end{equation}
and an \emph{update step} which integrates the information coming
from a new measurement $\boldsymbol{y}_{n+1}$ \emph{
\begin{equation}
P\left(\boldsymbol{x}_{n+1},t_{n+1}\mid\boldsymbol{y}_{1},t_{1};\cdots;\boldsymbol{y}_{n+1},t_{n+1}\right)=\frac{P\left(\boldsymbol{y}_{n+1}\mid\boldsymbol{x}_{n+1}\right)P\left(\boldsymbol{x}_{n+1},t_{n+1}\mid\boldsymbol{y}_{1},t_{1};\cdots;\boldsymbol{y}_{n},t_{n}\right)}{\int P\left(\boldsymbol{y}_{n+1}\mid\boldsymbol{x}_{n+1}\right)P\left(\boldsymbol{x}_{n+1},t_{n+1}\mid\boldsymbol{y}_{1},t_{1};\cdots;\boldsymbol{y}_{n},t_{n}\right)\diff\boldsymbol{x}_{n+1}}\label{eq:updateSTP}
\end{equation}
}

\end{widetext}which is an application of the Bayes' theorem. Our
point is that the information contained in $P\left(\boldsymbol{x},t\mid\boldsymbol{y}_{1},t_{1};\cdots;\boldsymbol{y}_{n},t_{n}\right)$
is larger than the one inside $P\left(\boldsymbol{x}\mid\boldsymbol{y}\right)$,
the last quantity being the only one that is used in the conventional
approach. Our hope is that with this increased information it will
be possible to improve the control strategy.

The complicated multidimensional integrals involved in the previous
equations can be computed explicitly only in very simple cases. For
example, when the probability distributions involved are gaussian
the previous steps lead to the Kalman filter~\citep{Kalman60,bozic1979digital,Maybeck:1990:KFI:93002.93293}.
Our locking strategy can be seen as inspired to some generalization
of the Kalman filter, as we will explain.

In Section~\ref{sec:The-model} we will introduce a simplified model
for the system, an oscillator subject to white noise, which will be
used to describe the locking problem in Section~\ref{sec:Locking}.
Though the system basically can be described by a gaussian distribution
in the state space, the problem of evaluating the locking performances
reduces to the solution of a Fokker Planck equation with absorbing
boundaries, which is not trivial and is studied numerically.

We draw our conclusions in Section~\ref{sec:Conclusions}. We propose
some natural generalizations, and in particular a method that can
be used to cope with poorly known parameters of the system, which
is based on the \emph{update step} of Eq.~\eqref{eq:updateSTP}.
The technique can in principle be used to allow the identification
of the suspension and the seismic noise model, but also of optical
parameters.

\section{\label{sec:The-model}The model}

The basic model we are interested to can be written in term of stochastic
motion equations which we write in the Ito form
\begin{eqnarray}
\diff V+2\gamma V\diff t+\omega_{0}^{2}\left(X\diff t-\tilde{\sigma}_{s}\diff W_{s}\right) & = & \mu^{-1}F_{ext}(t)dt\nonumber \\
\diff X & = & V\diff t\label{eq:emotion}
\end{eqnarray}
Here $X$ represents the variation of the length of the cavity with
respect to a reference position $L_{0}$ which is the equilibrium
one when the laser is switched off. This is an harmonic oscillator
of mass $\mu$ attached to a point which is subject to seismic motion
in presence of a viscous damping proportional to $\gamma$. The term
$F_{ext}$ is an external force which is introduced for future convenience. 

For the sake of simplicity the seismic motion is modeled as a white
noise of spectral variance $\tilde{\sigma}_{s}^{2}$, so $\diff W_{s}$
is a Wiener process, which is a quite crude approximation. A viscous
modelization of the damping can also be inaccurate in some cases. 

More accurate models can be cumbersome in the time domain, though
feasible. For example, in a suspended mirror dissipative effects are
typically described by adding a small imaginary part to the stiffness
constant $k$ in the frequency domain (structural damping). These
details would obscure the most relevant points we are interested to,
so we ignore them here, however in the final section (Subsection~\ref{sub:Improving-the-model})
we will discuss about a systematic way to introduce some improvement. 

A description of the cavity in the state space is easily obtained
by grouping the dynamic variables as $\boldsymbol{x}=X\boldsymbol{e}_{1}+\omega_{0}^{-1}V\boldsymbol{e}_{2}$.
The stochastic motion equations become
\begin{equation}
\diff\boldsymbol{x}=\boldsymbol{F}\left(\boldsymbol{x}\right)\diff t+\boldsymbol{\beta}\diff W\label{eq:stochemon}
\end{equation}
where (setting $f=\mu^{-1}\omega_{0}^{-1}F_{ext}$) 
\begin{eqnarray*}
\boldsymbol{F}\left(\boldsymbol{x}\right) & = & \left(\begin{array}{cc}
0 & \omega_{0}\\
-\omega_{0} & -2\gamma
\end{array}\right)\boldsymbol{x}+\left(\begin{array}{c}
0\\
f(t)
\end{array}\right)\equiv\mathbb{K}\boldsymbol{x}+f(t)\boldsymbol{e}_{2}
\end{eqnarray*}
and $\boldsymbol{\beta}=\omega_{0}\tilde{\sigma}_{s}\boldsymbol{e}_{2}$.
The evolution of the probability distribution $P\left(\boldsymbol{x},t\mid\boldsymbol{x}^{\prime},t^{\prime}\right)$
is obtained by solving the Fokker-Planck equation induced by~(\ref{eq:stochemon})
which reads
\begin{equation}
\frac{\partial P\left(\boldsymbol{x},t\mid\boldsymbol{x}^{\prime},t^{\prime}\right)}{\partial t}=-\boldsymbol{\nabla}_{\boldsymbol{x}}\cdot\boldsymbol{J}(x,t)\label{eq:FokkerPlanck}
\end{equation}
where 
\begin{equation}
\boldsymbol{J}(x,t)=\boldsymbol{F}\left(\boldsymbol{x}\right)P\left(\boldsymbol{x},t\mid\boldsymbol{x}^{\prime},t^{\prime}\right)-\frac{1}{2}\boldsymbol{\beta}\left(\boldsymbol{\beta}\cdot\boldsymbol{\nabla}_{\boldsymbol{x}}\right)P\left(\boldsymbol{x},t\mid\boldsymbol{x}^{\prime},t^{\prime}\right)\label{eq:ProbabilityCurrent}
\end{equation}
can be interpreted as a probability current. The first contribution
the the current in Equation~(\ref{eq:ProbabilityCurrent}) gives
a deterministic transport of the probability distribution, while the
second one generates a diffusion effect proportional to the noise.
Note that in our model the vector $\boldsymbol{\beta}$ has only the
second component different from zero, and as a consequence the same
is true also for the diffusion current. The reason is that the diffusion
is generated by the random seismic noise force, which can randomize
directly only the velocity. This randomization is next propagated
to the position variable by the deterministic dynamic.

The generating function $W_{S}$ of the (connected) momenta of $P$
over some region of interest $S$ is defined by
\[
e^{W_{S}}=\int_{S}e^{\boldsymbol{\eta}\cdot\boldsymbol{x}}P\left(\boldsymbol{x},t\mid\boldsymbol{x}^{\prime},t^{\prime}\right)d^{2}\boldsymbol{x}\equiv\left\langle e^{\boldsymbol{\eta}\cdot\boldsymbol{x}}\right\rangle _{S}
\]
and starting from the Fokker Planck equation~(\ref{eq:FokkerPlanck})
it can be shown that it is governed by the equation

\begin{eqnarray}
\frac{\partial W_{S}}{\partial t} & = & e^{-W_{S}}\boldsymbol{\eta}\cdot\boldsymbol{F}\left(\nabla_{\boldsymbol{\eta}}\right)e^{W_{S}}+\frac{1}{2}\left(\boldsymbol{\beta}\cdot\boldsymbol{\eta}\right)^{2}\label{eq:evolWs}
\end{eqnarray}

Here we neglected boundary terms which are not relevant if we are
interested in the evolution of $P$ over all the state space. By differentiating
one and two times and setting $\boldsymbol{\eta}=0$ we obtain the
motion equation for the expectation value of the state vector

\[
\frac{\partial}{\partial t}\left\langle \boldsymbol{x}\right\rangle =\left\langle \boldsymbol{F}\right\rangle 
\]
and the motion equation for its covariance array $\mathbb{C}\equiv\left\langle \boldsymbol{x}\otimes\boldsymbol{x}\right\rangle -\left\langle \boldsymbol{x}\right\rangle \otimes\left\langle \boldsymbol{x}\right\rangle $
which is
\[
\frac{\partial}{\partial t}\mathbb{C}=\left\langle \boldsymbol{x}\otimes\boldsymbol{F}\right\rangle +\left\langle \boldsymbol{F}\otimes\boldsymbol{x}\right\rangle -\left\langle \boldsymbol{x}\right\rangle \otimes\left\langle \boldsymbol{F}\right\rangle -\left\langle \boldsymbol{F}\right\rangle \otimes\left\langle \boldsymbol{x}\right\rangle +\boldsymbol{\beta}\otimes\boldsymbol{\beta}
\]
In our model $\boldsymbol{F}$ depends linearly on $\boldsymbol{x}$,
and we get
\begin{equation}
\frac{\partial}{\partial t}\left\langle \boldsymbol{x}\right\rangle =\mathbb{K}\left\langle \boldsymbol{x}\right\rangle +f(t)\boldsymbol{e}_{2}\label{eq:evolMU}
\end{equation}
which is exactly the equation of motion for the state without seismic
noise and 
\begin{equation}
\frac{\partial}{\partial t}\mathbb{C}=\mathbb{K}\mathbb{C}+\mathbb{C}\mathbb{K}^{T}+\boldsymbol{\beta}\otimes\boldsymbol{\beta}\label{eq:evolC}
\end{equation}
which is unaffected by the external force $f$. 

The explicit solution of these equations can be written using the
matrix $\mathbb{U}\left(t\right)\equiv e^{\mathbb{K}t}$ (see Eq.~\eqref{eq:expK}).
For the average value we get
\begin{equation}
\left\langle \boldsymbol{x}\left(t\right)\right\rangle =\mathbb{U}\left(t-t^{\prime}\right)\left\langle \boldsymbol{x}\left(t^{\prime}\right)\right\rangle +\int_{t^{\prime}}^{t}\mathbb{U}\left(t-\tau\right)\boldsymbol{e}_{2}\,f(\tau)d\tau\label{eq:evolaverage}
\end{equation}
When the external force is absent $\left\langle \boldsymbol{x}\right\rangle \rightarrow0$
on a time scale $\gamma^{-1}$, which is also the time scale in which
the initial condition is forgotten.

For the covariance array we obtain
\begin{equation}
\mathbb{C}\left(t\right)=\mathbb{U}\left(t-t^{\prime}\right)\mathbb{C}(t^{\prime})\mathbb{U}\left(t-t^{\prime}\right)^{T}+\mathbb{Q}\left(t-t^{\prime}\right)\label{eq:evolcovariance}
\end{equation}
with
\begin{eqnarray*}
\mathbb{Q}\left(t-t^{\prime}\right) & = & \int_{t^{\prime}}^{t}\mathbb{U}\left(t-\tau\right)\boldsymbol{\beta}\boldsymbol{\beta}^{T}\mathbb{U}\left(t-\tau\right)^{T}d\tau
\end{eqnarray*}
 The initial value of $\mathbb{C}$ goes to zero exponentially on
a timescale $(2\gamma)^{-1}$ while the noise induces the contribution
$\mathbb{Q}$ (see Equation~\eqref{eq:varianceQ} in the Appendix
for the explicit expression) which is initially zero and converges
on the same time scale to 
\[
\lim_{\tau\rightarrow\infty}\mathbb{Q}=\lim_{\tau\rightarrow\infty}\mathbb{C}=\left(\begin{array}{cc}
\sigma_{\infty}^{2} & 0\\
0 & \sigma_{\infty}^{2}
\end{array}\right)
\]
with
\[
\sigma_{\infty}^{2}\equiv\frac{\omega_{0}^{2}\tilde{\sigma}_{s}^{2}}{4\gamma}\equiv\frac{\tilde{\sigma}_{f}^{2}}{4\gamma\mu^{2}\omega_{0}^{2}}
\]
which correspond to two uncorrelated components of the state vector
with the same variance $\sigma_{\infty}^{2}$. The parameter $\tilde{\sigma}_{f}^{2}$
introduced is the spectral variance of a force equivalent to the seismic
displacement. 

The dynamic of our model preserve the gaussian character of a probability
distribution, owing to its linearity, and a generic gaussian solution
of the Fokker Planck equation can be written~\citep{mazo2002brownian}

\begin{equation}
P\left(\boldsymbol{x},t\mid\boldsymbol{x}^{\prime},t^{\prime}\right)={\cal N}\left(\boldsymbol{x};\overline{\boldsymbol{x}},\mathbb{C}\right)\label{eq:conditioned}
\end{equation}

By setting $\left\langle \boldsymbol{x}\left(t^{\prime}\right)\right\rangle =\boldsymbol{x}^{\prime}$
in Equation~(\ref{eq:evolaverage}), $\mathbb{C}\left(t^{\prime}\right)=0$
in Equation~(\ref{eq:evolcovariance}) and by substituting in Equation~(\ref{eq:conditioned})
we find in particular

\begin{widetext}

\begin{equation}
P\left(\boldsymbol{x},t\mid\boldsymbol{x}^{\prime},t^{\prime}\right)={\cal N}_{\boldsymbol{x}}\left(\mathbb{U}\left(t-t^{\prime}\right)\boldsymbol{x}^{\prime}+\int_{t^{\prime}}^{t}\mathbb{U}\left(t-\tau\right)\boldsymbol{e}_{2}\,f(\tau)d\tau,\mathbb{Q}\left(t-t^{\prime}\right)\right)\label{eq:conditioned-1}
\end{equation}

\end{widetext}which correspond to the initial condition $P\left(\boldsymbol{x},t^{\prime}\mid\boldsymbol{x}^{\prime},t^{\prime}\right)=\delta\left(\boldsymbol{x}-\boldsymbol{x}^{\prime}\right)$.

\section{\label{sec:Locking}Locking}

The amplitude of the field reflected by the cavity can be written
as $\Phi_{r}={\cal R}\Phi_{i}$. When $X$ changes slowly compared
with the bandwidth of the cavity, which is a very good approximation
in the cases we are interested to, we can write 
\begin{eqnarray*}
{\cal R} & = & \frac{r_{2}e^{2i\phi}-r_{1}}{1-r_{1}r_{2}e^{2i\phi}}\simeq\frac{r_{2}e^{2i\phi}-r_{1}}{1-r_{1}r_{2}}\frac{1}{1+\frac{{\cal {\cal F}}}{\pi}\left(1-e^{2i\phi}\right)}
\end{eqnarray*}
Here $\phi=2\pi\left(L_{0}+X\right)/\lambda_{\ell}$ is the phase
shift over the length of the cavity and $\lambda_{\ell}$ the wavelength
of the laser beam. We do not suppose that the cavity resonates at
$X=0$, so we write $L_{0}=L_{res}+L_{det}$ where the resonant length
$L_{res}$ is an integer multiple of $\lambda_{\ell}/2$ and $-1/4<L_{det}/\lambda_{\ell}<1/4$.
The reflectivities $r_{i}$, the transmissivities $t_{i}$ and the
losses of the mirrors are connected by the relation $r_{i}^{2}+t_{i}^{2}=1-p_{i}$~\footnote{Here and in the following we neglect the effect of optical losses.},
with $i=1,2$.

\begin{figure}
\includegraphics[width=1\columnwidth]{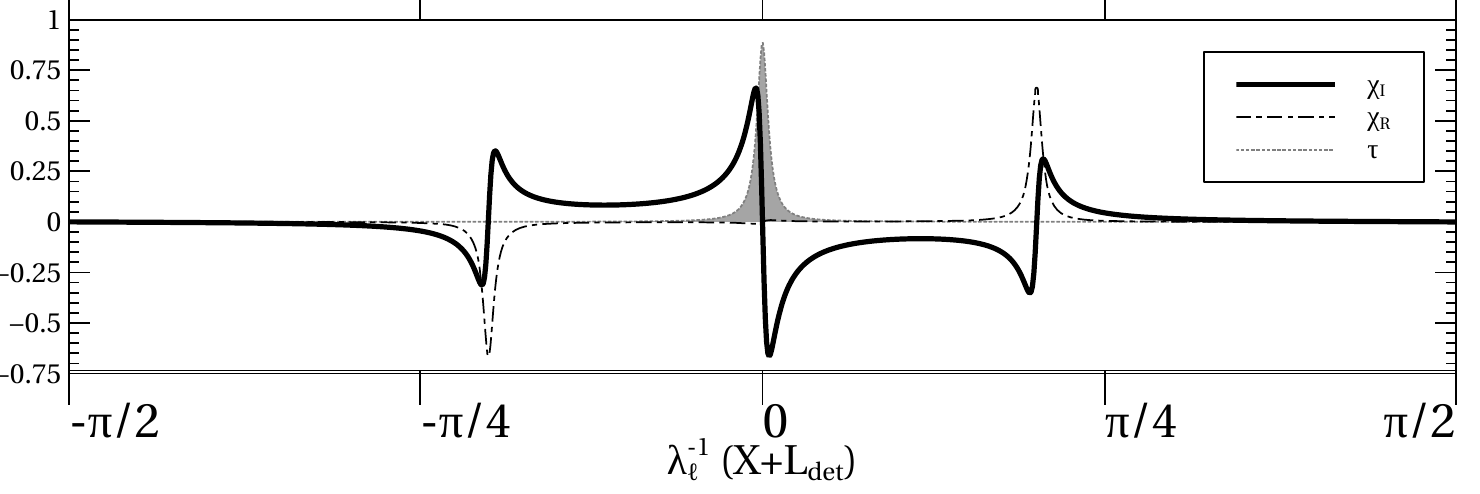}

\protect\caption{\label{fig:PDHsignal}The \protect\ac{PDH} signal $\chi$ for a resonant
cavity, as a function of the phase shift $\lambda_{\ell}^{-1}(X+L_{det})$.
In this particular case $r_{1}=0.99$, $r_{2}=0.98$ and $L_{0}=1\unit{m}$,
which gives a free spectral range of $f_{FSR}=150\unit{MHz}$. The
modulation frequency is $f_{m}=(2\pi)^{-1}\omega_{m}=30\unit{MHz}$.
The continuous line is the imaginary part $\chi_{I}=\mbox{Im}\,\chi$
of the \protect\ac{PDH} signal, the dotted line its real part $\chi_{R}=\mbox{Re}\,\chi$.
The dotted filled graph is the ratio $\tau$ between the transmitted
intensity and the input one. }
\end{figure}

The phase shift induced by the cavity can be measured with the \acl{PDH}
technique~\citep{PDH,PDHpedagogical}, which gives a complex signal
$\chi$ that can be written as 

\begin{equation}
\chi\equiv\chi_{R}+i\chi_{I}={\cal R}\left(\phi\right){\cal R}\left(\phi_{+}\right)^{*}-{\cal R}\left(\phi\right)^{*}{\cal R}\left(\phi_{-}\right)
\end{equation}
and $\phi_{\pm}=c^{-1}\left(\omega_{\ell}\pm\omega_{m}\right)\left(L_{0}+X\right)\simeq\phi\pm c^{-1}\omega_{m}L_{0}$
is the phase shift of the two \ac{PDH} sidebands over the cavity
length. An example of $\chi$ is given in Figure~\ref{fig:PDHsignal}
for a somewhat arbitrary choice of the relevant parameters. It is
evident that the connection between the cavity displacement and $\chi$
is linear only in a small interval $O\left(\lambda_{\ell}{\cal F}^{-1}\right)$
around the resonance, where it can be used as an error signal. 

Another useful signal is given by the ratio between the transmitted
intensity and the input one, $\tau\equiv\left|\Phi_{t}\right|^{2}/\left|\Phi_{i}\right|^{2}$,
where $\Phi_{t}={\cal T}\Phi_{i}$ and
\begin{equation}
{\cal T}=\frac{t_{1}t_{2}e^{i\phi}}{1-r_{1}r_{2}e^{2i\phi}}\simeq\frac{t_{1}t_{2}e^{i\phi}}{1-r_{1}r_{2}}\frac{1}{1+\frac{{\cal {\cal F}}}{\pi}\left(1-e^{2i\phi}\right)}\Phi_{i}\label{eq:transmitted}
\end{equation}
In order for $\tau$ to be different from zero we must allow for a
(small) transmissivity $t_{2}$. This is also plotted in Figure~\ref{fig:PDHsignal}:
note that it can be used as a trigger, as it becomes different from
zero in a significant way only in the linear region of the \ac{PDH}
signal.
\begin{figure}
\includegraphics[width=1\columnwidth]{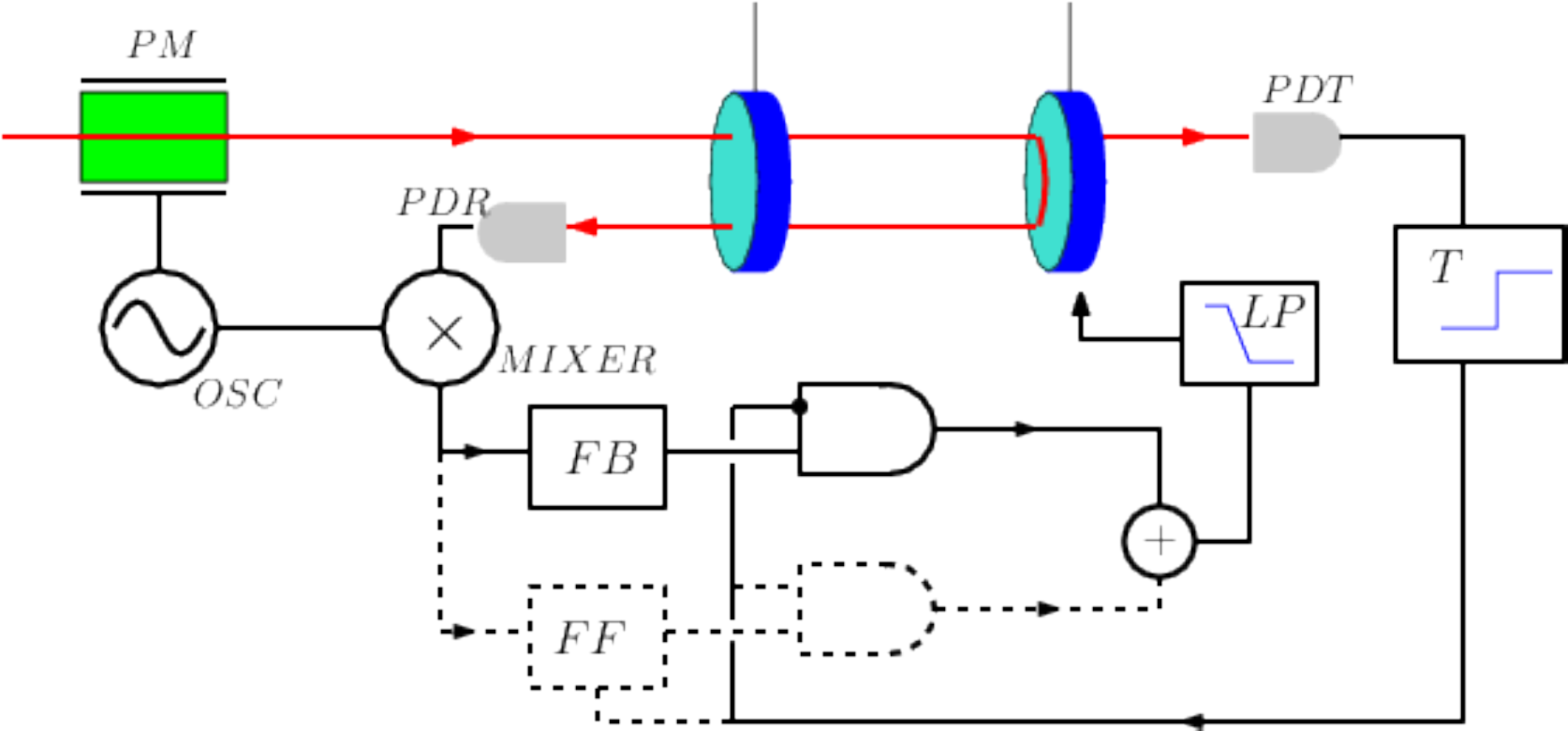}

\protect\caption{\label{fig:Scheme}A schematic representation of the locking procedure.
Both the light transmitted by the cavity, measured by the photodiode
$PDT$, and the phase shift Pound Drever signal measured by the photodiode
$PDR$ are used. The transmitted signal is used as a trigger, to discriminate
between the inside \protect\ac{LRAR} and the outside \protect\ac{LRAR}
regimes. Inside the \protect\ac{LRAR} a standard feedback force $FB$
is applied to the system, while outside the feed forward force $FF$
is used. The feed forward force is provided by $FF$ taking into account
the value of the cavity velocity at the exit of \protect\ac{LRAR}.
The low pass filter allow for a smooth transition between feedback
and feed forward.}
\end{figure}

In Figure~\ref{fig:Scheme} we schematized the proposed improved
locking strategy. The non dashed parts represent the usual locking
scheme, the dashed one our additions. The two parts are activated
respectively inside and outside the \ac{LRAR}: in the first case
we have a feed backward scheme, in the second a feed forward one. 

We can assume that, up to a negligible measurement error, the state
of the system will be known exiting the \ac{LRAR}. The main problem
for locking acquisition is the value of the velocity: if too large,
the usual lock attempt will require a large feedback force. Our strategy
can now be described as follows: when the cavity exit the \ac{LRAR}
we will stop to measure its state. But, using our knowledge of the
probability distribution during the out of resonance period, we apply
a feed forward force $f$ optimized in such a way to reduce the cavity
speed, when it will reenter the \ac{LRAR}. 

In particular, suppose the exiting velocity of the cavity to be $v_{exit}>0$.
The cavity can reenter the \ac{LRAR} around the same resonance, with
a final velocity $v<0$, or around the first one on the right, with
a final velocity $v>0$. A key quantity will be the \ac{LRAR} reentering
velocity distribution $P_{RVD}\left(v;v_{exit}^{2},f\right)$, which
is a function of $v_{exit}^{2}$ and a functional of $f$. Starting
from $P_{RVD}\left(v;v_{exit}^{2},f\right)$ it will easy to evaluate
the square several parameter of interest.

\begin{figure}
\includegraphics[width=1\columnwidth]{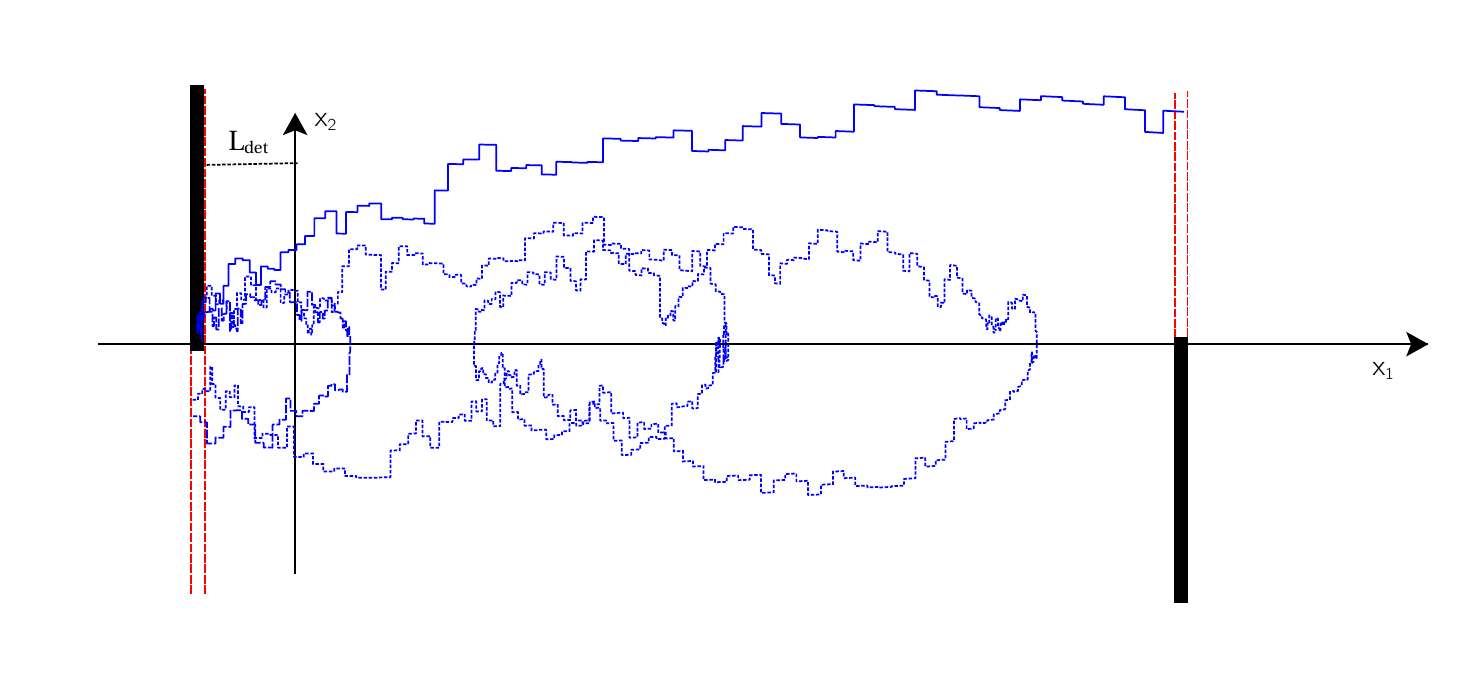}

\protect\caption{\label{fig:Trajectories}A sample of schematic trajectories which
contribute to the \protect\ac{LRAR} reentering velocity distribution
$P_{RVD}$, with the same initial velocity. The continuous line contributes
to the $v>0$ part of $P_{RVD}$, the dashed and dotted lines to the
$v<0$ one. Note that the noise allows the cavity to invert several
time its speed before reentering in the \protect\ac{LRAR} (dotted
line). For illustrative purposes the trajectories are discretized,
and have been simulated as a sequence of a vertical step due to the
noise and a transport one. The tick lines are the absorbing boundaries,
where $P\left(\boldsymbol{x},t\mid\boldsymbol{x}_{0},t_{0}\right)=0$.}
\end{figure}

In order to calculate $P_{RVD}\left(v;v_{exit}^{2},f\right)$ we need
to solve the Fokker Planck equation with peculiar boundary conditions
which represent absorbing barriers at the \ac{LRAR}s. In Figure~\ref{fig:Trajectories}
we represented the strip of the state space between two \ac{LRAR}
of interest (double dashed lines) centered in $x_{1}\equiv x_{L}$
and $x_{1}\equiv x_{R}=x_{L}+\frac{\lambda_{\ell}}{2}$. When the
state of the cavity enters a \ac{LRAR} we want to stop to consider
it, so we set $P\left(\boldsymbol{x},t\mid\boldsymbol{x}^{\prime},t^{\prime}\right)=0$
outside the strip. In our particular case there is not a diffusion
current in the horizontal direction: this means that $P$ needs not
to be continuous across the vertical boundaries of the strip. However
the probability current on the boundaries
\begin{align*}
J_{1}\left(\boldsymbol{x},t\right) & =\left.P\left(\boldsymbol{x},t\mid\boldsymbol{x}^{\prime},t^{\prime}\right)\boldsymbol{F}\left(\boldsymbol{x}\right)\cdot\boldsymbol{e}_{1}\right|_{x_{1}=x_{L,R}}\\
 & =\left.\omega_{0}P\left(\boldsymbol{x},t\mid\boldsymbol{x}^{\prime},t^{\prime}\right)x_{2}\right|_{x_{1}=x_{L,R}}
\end{align*}
must be directed outward, and this gives the boundary conditions,
valid for $t>t^{\prime}$,
\[
\left.P\left(\boldsymbol{x},t\mid\boldsymbol{x}^{\prime},t^{\prime}\right)x_{2}\right|_{x_{1}=x_{L},x_{2}>0}=0
\]
and

\[
\left.P\left(\boldsymbol{x},t\mid\boldsymbol{x}^{\prime},t^{\prime}\right)x_{2}\right|_{x_{1}=x_{R},x_{2}<0}=0
\]
which are represented by the black thick lines in Figure~\ref{fig:Trajectories}.

After finding the solution $P^{*}$ which satisfy $P^{*}\left(\boldsymbol{x},t^{\prime}\mid\boldsymbol{x}^{\prime},t^{\prime}\right)=\delta\left(\boldsymbol{x}-\boldsymbol{x}^{\prime}\right)$
with $\boldsymbol{x}^{\prime}=x_{L}\boldsymbol{e}_{1}+\omega_{0}^{-1}v_{exit}\boldsymbol{e}_{2}$
we can calculate the positive velocity part of $P_{RVD}$ as
\begin{multline}
P_{RVD}\left(v;v_{exit}^{2},f\right)=\left.\int_{0}^{\infty}J_{1}\right|_{x_{1}=x_{R},x_{2}=\omega_{0}^{-1}v}dt\\
=v\int_{0}^{\infty}P^{*}\left[\left(\begin{array}{c}
x_{R}\\
\omega_{0}^{-1}v
\end{array}\right),t;\left(\begin{array}{c}
x_{L}\\
\omega_{0}^{-1}v_{exit}
\end{array}\right),0\right]dt\label{eq:PRVDplus}
\end{multline}

and the negative velocity part as
\begin{multline}
P_{RVD}\left(v;v_{exit}^{2},f\right)=\left.\int_{0}^{\infty}J_{1}\right|_{x_{1}=x_{L},x_{2}=\omega_{0}^{-1}v}dt\\
=v\int_{0}^{\infty}P^{*}\left[\left(\begin{array}{c}
x_{L}\\
\omega_{0}^{-1}v
\end{array}\right),t;\left(\begin{array}{c}
x_{L}\\
\omega_{0}^{-1}v_{exit}
\end{array}\right),0\right]dt\label{eq:PRVDminus}
\end{multline}

As a consequence of the absorbing boundaries $P^{*}$ is not a gaussian
distribution, and its normalization is not conserved: its integral
over the strip at the time $t>t^{\prime}$ gives the probability for
the cavity of not being reentered the \ac{LRAR} at that time. 

Supposing that we do not care about the particular \ac{LRAR} where
the system will be locked, we can design our feed forward force in
such a way to maximize square velocity reduction probability 
\[
P_{red}\left(v_{exit}^{2},f\right)=\int_{-v_{exit}}^{v_{exit}}P_{RVD}\left(v;v_{exit}^{2},f\right)dv
\]
In the best case we could obtain $P_{red}\left(v_{exit}^{2},f\right)>1/2\,\forall v_{exit}^{2}$:
this would guarantees the possibility of ``cooling'' systematically
the cavity until the locking becomes possible.

The maximization must be done over a class of functions $f$ which
satisfy some set of requirements: $f$ must not be too large and with
not a too large frequency bandwidth. We will write $f$ as
\[
f(t)=\int_{-\infty}^{+\infty}K\left(t-t^{\prime}\right)f_{0}(t^{\prime})dt^{\prime}
\]
where $\left|f_{0}(t)\right|<F_{0}$ and $K(t)$ represent a linear,
time invariant, causal low pass filter. 

Owing to the non trivial boundary conditions, we do not attempt here
to find an analytical solution of the Fokker Planck equation, neither
to determine the optimal control. Instead we evaluate numerically
the performances of a set of simple control strategies:
\begin{description}
\item [{Strategy~1}] \label{enu:FirstOption}When $v_{exit}$ is large,
the only option is to try to slow down the cavity until it enters
the next \ac{LRAR} on the right. A possible strategy is to apply
the most negative constant force available $f_{0}=-F_{0}$. The real
force will be given by 
\begin{equation}
f(t)=-F_{0}g(t)\label{eq:strategy1}
\end{equation}
where $g(t)=\int_{0}^{+\infty}K(t-t^{\prime})dt^{\prime}$ is the
unit step response of the filter described by $K(t)$. There are no
free parameters.
\item [{Strategy~2}] A second possibility, which should be effective in
the intermediate region for $v_{exit}$, is to initially accelerate
the cavity and then to decelerate it. This would make possible to
bring the cavity to the \ac{LRAR} on the right in a short time (reducing
the effects of the diffusion) with a small final velocity. In this
case the applied force will be
\begin{equation}
f(t)=F_{0}\left[g(t)-2g\left(t-\tau_{1}\right)\right]\label{eq:strategy2}
\end{equation}
 with a free parameter $\tau_{1}$ to adjust.
\item [{Strategy~3}] With the third strategy we attempt to bring back
the cavity to its starting \ac{LRAR}. This is expected to be effective
for low enough values of $v_{exit}$. We need a deceleration phase
followed by an acceleration one, needed to stop the typical cavity
which is moving back. This gives 
\begin{equation}
f(t)=-F_{0}\left[g(t)-2g(t-\tau_{1})\right]\label{eq:strategy3}
\end{equation}
and also in this case there is a single adjustable parameter. Note
that the option~\ref{enu:FirstOption} can be seen as the $\tau_{1}\rightarrow\infty$
limit of this one. 
\end{description}

\subsection{\label{sub:Numerical-results}Numerical results}

To be definite we choose a set of parameters which are somewhat representative
of the typical scenario one encounter in an interferometric gravitational
wave detector, setting $\omega_{0}=2\pi\,\unit{rad\,s^{-1}}$, $\gamma\simeq10^{-3}\omega_{0}$
and $\lambda_{\ell}=10^{-6}\unit{m}$. Setting an upper frequency
cut-off around $f_{c}\simeq100\unit{Hz}$ for both seismic noise and
control force bandwidth we have also $\tilde{\sigma}_{s}=10^{-6}/\sqrt{f_{c}}\simeq10^{-7}\unit{m\,Hz^{-1/2}}$
(supposing the root mean square of seismic displacement to be $10^{-6}\unit{m}$).
We suppose the maximal force we can apply to the mirror to be of the
order of $10^{-3}\unit{N}$, which for a cavity mass\footnote{The cavity mass $\mu$ is the reduced mass of its two mirrors.}
of $\mu=20\unit{kg}$ gives $F_{0}=8\times10^{-6}\unit{ms^{-1}}$.

It will be useful to give some order of magnitude estimates in absence
of external forces. With the given parameters  the length of the free
cavity is spread over $\sigma_{\infty}\simeq10^{-5}\unit{m}$, and
its typical velocity is $\omega_{0}\sigma_{\infty}\simeq6\times10^{-5}\unit{ms^{-1}}$,
which gives a typical time needed to move from a \ac{LRAR} to the
next one of $\tau_{T}\simeq4\times10^{-2}\unit{s}$. 

Note that both $\omega_{0}\tau_{T}$ and $\gamma\tau_{T}$ are small,
so we expect the details of the dynamics to be relevant only for for
velocities much smaller than the typical one. In the same typical
regime the relative spread of the initial velocity can be approximated
as $\sigma_{v}/(\omega_{0}\sigma_{\infty})\simeq2\sqrt{\gamma\tau_{T}}$:
once again we expect diffusion effects connected to the noise to be
relevant only for velocities much smaller than the typical one. 

This means that when $v_{exit}$ has a typical value, or larger, we
can estimate the result of the feed forward force looking only at
the average value of the probability distribution. In this case by
applying the first strategy we obtain always a reduction of the cavity
velocity. If we have an infinitely large frequency bandwidth at our
disposal then $f(t)=-F_{0}$ and the final velocity will be 
\[
v^{\prime}\simeq\sqrt{v_{exit}^{2}-F_{0}\lambda_{\ell}\omega_{0}}\simeq v_{exit}\left[1-\left(\frac{0.5\times10^{-6}}{v_{exit}}\right)^{2}\right]
\]
In the real case we should take into account the effect of the low
pass filter impulse response $K(t)$ in Equation~(\ref{eq:strategy1}):
the force will need a time $O(\omega_{c}^{-1})$ to rise to the largest
value available, and the velocity reduction will be suppressed if
$\omega_{c}\tau_{T}$ is small. In short, we are sure to obtain a
high $P_{red}$, the velocity reduction can be small, but obviously
always larger than the one obtainable in the \ac{LRAR} as $\tau_{T}$
is larger compared with the time spent inside the \ac{LRAR} by a
factor ${\cal F}$.

What happens when $v_{exit}$ is smaller than $\omega_{0}\sigma_{\infty}$
is less obvious. In principle diffusion effect in the phase space
due to the seismic noise could hamper the possibility of reducing
the cavity speed. 

We evaluated $P_{RVD}\left(v;v_{exit}^{2},f\right)$ by repeatedly
integrating numerically the motion equations~(\ref{eq:stochemon})
with the appropriate initial position and velocity, using a simple
leap frog scheme. This gives an ensemble of trajectories in the phase
space (some of them are represented in Figure~\ref{fig:Trajectories}).
When they reach the absorbing boundaries the integration is stopped
and the final velocity is stored until a sufficient statistic is obtained. 

\begin{figure}
\includegraphics[width=1\columnwidth]{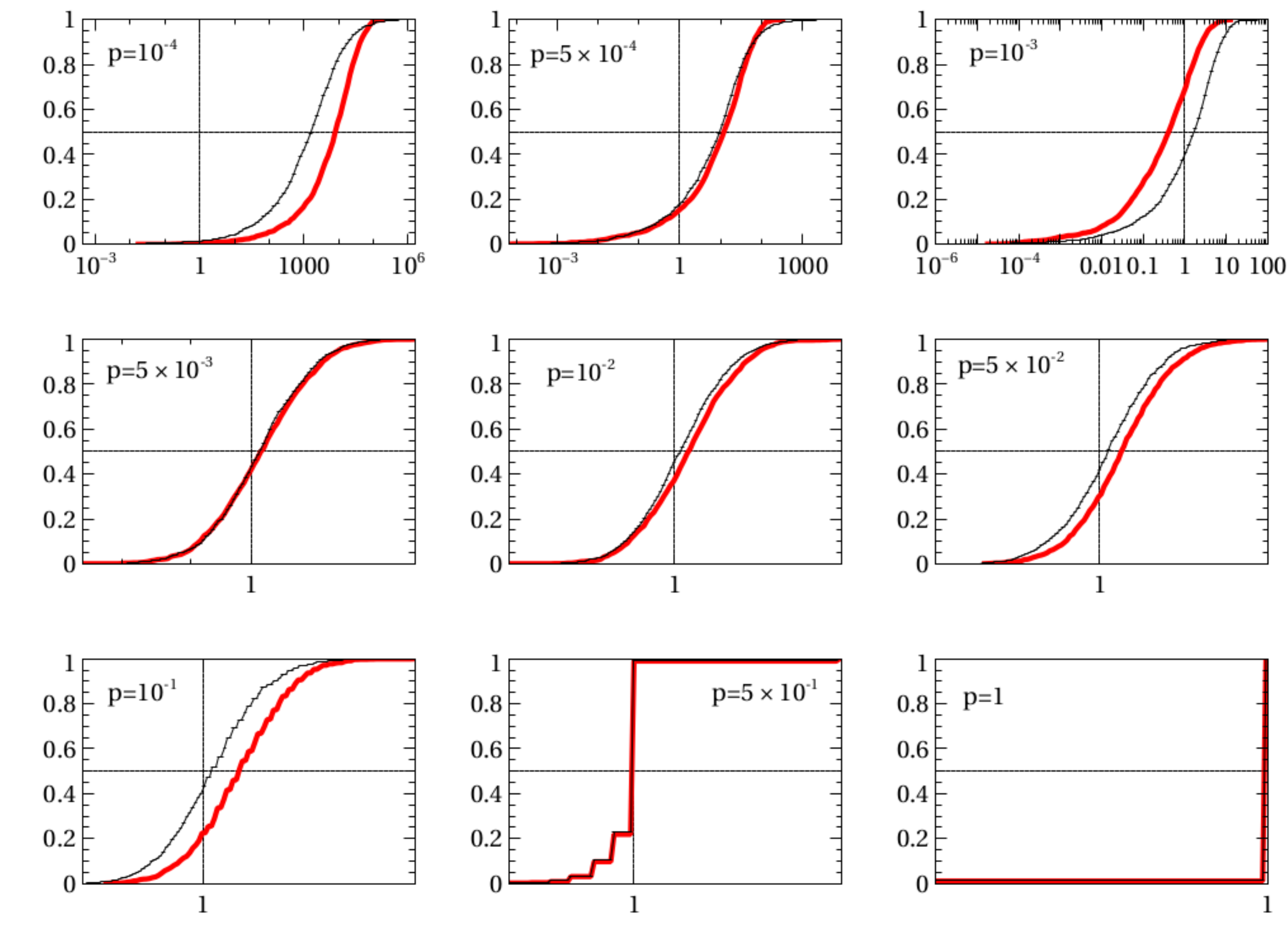}

\protect\caption{\label{fig:Strategy1}Results for the cumulative probability distribution
of the final square velocity when the first control strategy is used,
for selected values of $v_{exit}=p\omega_{0}\sigma_{\infty}$ and
$F_{0}=8\times10^{-6}\unit{ms^{-1}}$ (bold lines). For comparison
the results with $F_{0}=0$ are also plotted (dotted lines). The value
of the square velocity on the horizontal axis is normalized to the
initial one. }
\end{figure}

In Figure~\ref{fig:Strategy1} the results for the first strategy
are reported. We plot the cumulative probability distribution of the
ratio $\left(v/v_{exit}\right)^{2}$, for a given $v_{exit}$ which
is given as a fraction of the typical one. The thick line correspond
to the distribution obtained when the control force is present. For
comparison the results without control force is also showed (thin
line).

The value of $F_{0}$ is keep constant, and we applied a third order
Butterworth low pass filter. By looking at the value of the cumulative
distribution when $\left(v/v_{exit}\right)^{2}=1$ we can see that
we are able to obtain a velocity reduction probability larger than
$1/2$ when $p=10^{-3}$.

When $p=5\times10^{-1}$ and $p=1$ we can't appreciate the result,
because as anticipated the final velocities are almost unchanged compared
with the initial one. When $p=10^{-4}$ our feedback seems to obtain
a results which is the opposite of the desired one. This can be understood
because in this case the cavity has a velocity which is very small
compared with the typical one: the noise fluctuations accelerate it,
and when we apply the control force there is a larger time available
for this acceleration.

The general conclusion is that the first strategy seems to work in
a specific range for $v_{exit}$. We did not attempt a full optimization
at this stage.

\begin{figure}
\includegraphics[width=1\columnwidth]{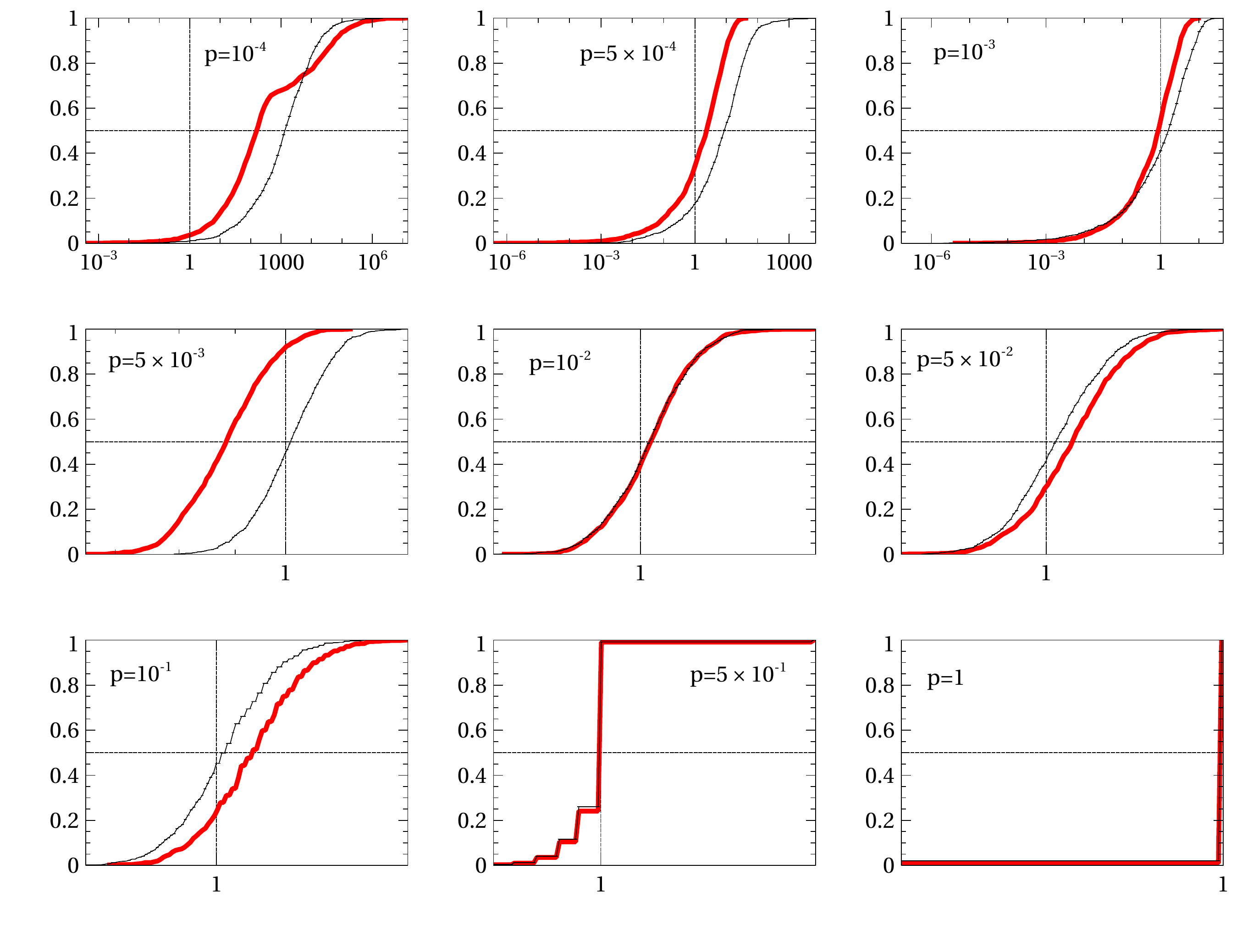}

\protect\caption{\label{fig:Strategy3}Results for the cumulative probability distribution
of the final square velocity when the third control strategy (with
$\tau_{1}=0.04\unit{s}$) is used, for selected values of $v_{exit}=p\omega_{0}\sigma_{\infty}$
and $F_{0}=8\times10^{-6}\unit{ms^{-1}}$ (bold lines). For comparison
the results with $F_{0}=0$ are also plotted (dotted lines). The value
of the square velocity on the horizontal axis is normalized to the
initial one. }
\end{figure}

In Figure~\ref{fig:Strategy3} similar results are reported for the
third strategy. We fixed a value for the free parameter $\tau_{1}$
which should be the optimal one in absence of fluctuations for $p=3\times10^{-3}$.
We see that we obtain the desired objective for $p=10^{-3}$ and $p=5\times10^{-3}$,
which is in agreement with the expectations. We did not attempt a
full investigation in this case neither, but we expect that with an
appropriate tuning of $\tau_{1}$, and with an appropriate use of
the second strategy when needed, it should be possible to slow down
the cavity whatever its initial velocity will be.

\section{\label{sec:Conclusions}Conclusions and perspectives}

We proposed a simple generalization of the common scheme used for
the lock acquisition, giving some initial numerical evidence that
the generalized scheme could get better performances. The discussed
strategy is ``blind'', in the sense that the additional control
force applied to the cavity when it is outside the resonance region
is a feed forward one, designed using only the last known value of
the state variables and the information about the cavity dynamics.
Several details have been neglected in this paper: we aimed only to
discuss the basic principles, and a detailed experimental understanding
is needed to discover potentially weak points.

There are however a couple of improvements that can be foreseen, and
will be the object of further investigation.

\subsection{\label{sub:Improving-the-model}Improving the model}

Our model of the cavity dynamics is quite simple. In a real situation,
the mirrors are suspended to a complex attenuation system needed to
reduce external seismic noise at the desired level. This means that
the simple oscillator considered in this paper should be substituted
by a chain of coupled ones. And, as mentioned initially, the model
for the dissipation is a rough one.

In a similar way, we modeled the seismic noise in the frequency band
of interest as white noise, while in a realistic scenario it will
have non trivial spectral peculiarities, namely it will be a colored
gaussian noise process.

We do not expect these neglected details to have a big impact on the
results. As a matter of fact, during its permanence between the small
region between two \ac{LRAR}s it will be quite a good approximation
to neglect completely the dependence of the mechanical force from
the position. 

The specific dissipation model can have a larger impact: in the viscous
case dissipation effects will be larger at higher velocity compared
with structural ones, so the estimation of the efficiency of the feed
forward strategy can be different. And in principle strong spectral
peculiarities of the seismic noise, introducing time correlations,
can make some difference. 

All these are modelization issues, that can lead to the introduction
of some unknown parameters. We stress that also in our simplified
model there are some parameters which are totally unknown (such as
$L_{det}$) or known with some uncertainty ($\gamma$, $\omega_{0}$).

In our initial discussion we introduced the \emph{prediction step}
described by Equation~(\ref{eq:predictionSTP}) and the \emph{update
step} described by Equation~(\ref{eq:updateSTP}). In designing the
feed forward scheme we used the \emph{prediction step} only, but the
\emph{update} one play an important role when we need to cope with
some unknown or partially known parameters $\boldsymbol{p}$. The
basic idea is to redefine the state variables writing
\[
\boldsymbol{x}^{ext}=\left(\begin{array}{c}
\boldsymbol{x}\\
\boldsymbol{p}
\end{array}\right)
\]
where the dimension of $\boldsymbol{x}$ can be larger than two to
accommodate a more refined mechanical model (a suspension chain, a
realistic damping mechanism). The motion equation for $\boldsymbol{x}$
will depend parametrically by the variables $\boldsymbol{p}$: now
we can consider the extended probability distribution $P\left(\boldsymbol{x},\boldsymbol{p},t\mid\cdots\right)$
and write a Fokker Planck equation for it.

A first possibility is to impose a trivial dynamic for $\boldsymbol{p}$
(no evolution at all) and to start from a given prior which describes
our ignorance of the $\boldsymbol{p}$'s values. After each \emph{prediction
step} the evolved $P\left(\boldsymbol{x},\boldsymbol{p},t\mid\cdots\right)$
will be compared with a new measurement during the \emph{update step},
which will select the regions of the probability space which better
represent the real system. 

If needed we can add a diffusive dynamics for the variables $\boldsymbol{p}$,
with the possibility of adapting to slow drifts of the system. We
mention that some of the parameters $\boldsymbol{p}^{seism}\in\boldsymbol{p}$
can be used to describe colored seismic noise $dW^{seism}$. This
can be done by writing
\[
dW^{seism}(t)=\int F\left(t-t^{\prime}\right)dW(t^{\prime})
\]
where $F(t)$ is some parametrized filter function and $dW$ a Wiener
process. Once again we can model our knowledge of seismic noise with
some prior, and we can introduce a drift for $\boldsymbol{p}^{seism}$
to allow the model to adapt.

In our model we completely neglect the effect of radiation pressure.
This is not a too bad approximation for our purposes, because radiation
pressure effects are depressed outside the \ac{LRAR}. When the radiation
pressure is large we expect however residual effect at the \ac{LRAR}
boundary, which can have an impact on the feed forward design. 

There are no problems in principle in introducing the radiation pressure
in our model. An important difference is that there will not be a
gaussian solution for the evolution of the cavity, as the equation
of motion will be nonlinear. This is not a great complication, because
also in the simplified model studied in this paper the probability
distribution of interest is not gaussian, owing to the absorbing boundaries. 

We mention that in principle it could be possible to compensate radiation
pressure effects by looking at the transmitted signal $\tau$, which
is just proportional to the laser intensity inside the cavity.

\subsection{\label{sub:Improving-the-locking}Improving the locking strategy}

A further improvement in the locking procedure performances could
be obtained unblinding (at least partially) the control strategy outside
the \ac{LRAR} region. This would convert our feed forward procedure
in a feedback one.

A possibility worth to be studied is the one of using the Kalman signal
$\chi$ (and the transmission signal $\tau$) in the nonlinear region.
Here the problem is not the nonlinear dependence of $\chi$ from the
cavity position, but its non univocity. During the \emph{update step}
this leads to the generation of a non gaussian probability distribution.
This can be parametrized with good accuracy as a gaussian misture,
and we are led to the concept of\emph{ particle filters}~\citep{Gustafsson2002,Pitt2012},
which can be seen as a way to parametrize a generic (not necessarily
gaussian) probability distribution in the state space. This has the
advantage of a simple implementation which does not requires tricky
linearizations, but comes at the expense of a larger computational
cost. 

We are currently investigating all these issues, and we plan to report
on them in a future paper.

\appendix

\section{Explicit expression of evolution operators}

We list for reference the explicit expressions of the operators which
appears in the evolution equations for our system. The fundamental
quantity is the exponential of the matrix $\mathbb{K}$
\begin{equation}
\mathbb{U}\left(t\right)\equiv e^{\mathbb{K}t}=e^{-\gamma t}\left(\begin{array}{cc}
\cos\Omega t+\frac{\gamma}{\Omega}\sin\Omega t & \frac{\omega_{0}}{\Omega}\sin\Omega t\\
-\frac{\omega_{0}}{\Omega}\sin\Omega t & \cos\Omega t-\frac{\gamma}{\Omega}\sin\Omega t
\end{array}\right)\label{eq:expK}
\end{equation}
where $\Omega^{2}=\omega_{0}^{2}-\gamma^{2}$. The inhomogeneous term
of the covariance matrix is given by

\begin{equation}
\mathbb{Q}(\tau)=\sigma_{\infty}^{2}\left[\left(1-e^{-2\gamma\tau}\right)\mathbb{I}-\frac{2\gamma}{\Omega}e^{-2\gamma\tau}\sin\Omega t\,\mathbb{M}(t)\right]\label{eq:varianceQ}
\end{equation}
where 
\[
\mathbb{M}(t)=\left(\begin{array}{cc}
\frac{\gamma}{\Omega}\sin\Omega t+\cos\Omega t & \frac{\omega_{0}}{\Omega}\sin\Omega t\\
\frac{\omega_{0}}{\Omega}\sin\Omega t & \frac{\gamma}{\Omega}\sin\Omega t-\cos\Omega t
\end{array}\right)
\]
and $\mathbb{I}$ is the identity matrix.

The explicit expression of a multivariate gaussian distribution for
the variable $\boldsymbol{x}$, with mean $\overline{\boldsymbol{x}}$
and covariance matrix $\mathbb{C}$ is given by

\begin{equation}
{\cal N}\left(\boldsymbol{x};\overline{\boldsymbol{x}},\mathbb{C}\right)=\frac{1}{2\pi\sqrt{\det\mathbb{C}}}\exp\left[-\frac{1}{2}\left(\boldsymbol{x}-\overline{\boldsymbol{x}}\right)^{T}\mathbb{C}^{-1}\left(\boldsymbol{x}-\overline{\boldsymbol{x}}\right)\right]\label{eq:conditioned-2}
\end{equation}

\section{List of acronyms used}

\begin{acronym}
\acro{PD}{Probability Distribution}\acro{PDH}{Pound Drever Hall}\acro{LRAR}{Linear Region Around Resonance}
\end{acronym}

\end{document}